\documentclass[jcp,twocolumn,groupedaddress,preprintnumbers,amsmath,amssymb,4paper,superscriptaddress,floatfix,twoside]{revtex4-1}

\usepackage[yyyymmdd,hhmmss]{datetime}

\usepackage[pdftex]{graphicx}
\usepackage{xcolor}
\usepackage{dcolumn}
\usepackage{bm}
\usepackage{amsmath}

\usepackage{natbib}
\usepackage[colorlinks=true,urlcolor=blue,citecolor=blue,linkcolor=blue,breaklinks=true]{hyperref}

\usepackage{orcidlink}

\usepackage{times}

\begin{document}

\sloppy 

\title{An OrthoBoXY-Method for Various Alternative Box Geometries}

\author{Johanna Busch\orcidlink{0000-0003-3784-0188}}
\affiliation{Institut f\"ur Chemie, Abteilung Physikalische und Theoretische Chemie, 
Universit\"at Rostock, Albert-Einstein-Str.\ 27, D-18059 Rostock, Germany}

\author{Dietmar Paschek\orcidlink{0000-0002-0342-324X}}
\email{dietmar.paschek@uni-rostock.de}
\affiliation{Institut f\"ur Chemie, Abteilung Physikalische und Theoretische Chemie, 
Universit\"at Rostock, Albert-Einstein-Str.\ 27, D-18059 Rostock, Germany}

\date{\today~at~\currenttime}

\begin{abstract}
We have shown in a recent contribution
\href{https://doi.org/10.1021/acs.jpcb.3c04492}{[{\em J.~Phys.~Chem.B} \textbf{127}, 7983-7987 (2023)]}
that for 
molecular dynamics (MD)
 simulations 
of isotropic fluids based on 
orthorhombic
periodic boundary conditions
with ``magic'' box length ratios of $L_z/L_x\!=\!L_z/L_y\!=\!2.7933596497$, the computed self-diffusion coefficients $D_x$ and $D_y$ in $x$- and $y$-direction 
become system size independent. They thus
represent the true self-diffusion coefficient $D_0\!=\!(D_x+D_y)/2$,
while the shear viscosity can 
be determined from diffusion 
coefficients in $x$-, $y$-, and $z$-direction, using the expression
$\eta\!=\!k_\mathrm{B}T\cdot 8.1711245653/[3\pi L_z(D_{x}+D_{y}-2D_z)]$. 
Here we present a more generalized version of this
``OrthoBoXY''-approach, which can be applied to any orthorhombic
MD box. We would like to test, whether it is possible to
improve the efficiency of the approach by
using a shape more akin to the cubic form, albeit with different
box-length ratios $L_x/L_z\!\neq\! L_y/L_z$ 
and $L_x\!<\!L_y\!<\!L_z$. We use simulations of 
systems of 1536 TIP4P/2005 water molecules 
as a benchmark
and explore different
box-geometries to determine the influence of the
box shape on the computed statistical uncertainties
for $D_0$ and $\eta$. 
Moreover, another ``magical'' set of box-length ratios is
discovered
with $L_y/L_z\!=\!0.57804765578$ and
$L_x/L_z\!=\!0.33413909235$, where the self-diffusion coefficient
in $x$-direction becomes system size independent,
such that $D_0\!=\!D_x$.
\end{abstract}

\maketitle

\section{Introduction}

The viscosity of a fluid and the diffusion coefficients of
its constituents provide us with
an important reference
for the understanding of a large variety of transport-related
phenomena.\cite{taylor_krishna,viscosity_handbook}
To investigate molecular transport properties, molecular dynamics (MD) 
simulations have demonstrated to be an
important novel source 
for delivering insights and
for providing reference data.\cite{maginn_2019}
For example,
MD simulations
can produce accurate reference data for systems, which are
otherwise difficult to measure, such as the 
multicomponent diffusion inside of
nanoporous materials,\cite{krishna_2009} or they are
enabling us to study conditions, which are experimentally difficult to replicate, 
such as the pressures and temperatures found
in the earth's interior.\cite{li_2022} 

However, self-diffusion coefficients within liquids obtained
from from MD simulations 
with periodic boundary conditions (PBCs)
can exhibit 
  a quite substantial
   system size dependence.\cite{celebi_2021,erdoes_2020,duenweg_1993,yeh_2004}
A review article by Celebi et al. provides a good overview 
of this topic.\cite{celebi_2021}
This effect is
caused by the altered
hydrodynamic interactions 
between particles in a periodic system,
and leads to an $L^{-1}$-dependence of the self-diffusion coefficients,
where $L$ represents the length of the cubic simulation box.\cite{duenweg_1993,yeh_2004,kikugawa_2015,voegele_2016,moultos_2016}
This behavior has been quantitatively analyzed for 
simulations of 
polymers in solution \cite{duenweg_1993}, 
TIP3P model water molecules, 
and Lennard-Jones particles \cite{yeh_2004}, 
as well as carbon dioxide, n-alkanes, and poly(ethylene glycol) dimethyl ethers 
for a wide variety of conditions.\cite{moultos_2016}
In their seminal contribution, Yeh and Hummer have shown that
determining {\em true} system size independent self-diffusion coefficients 
thus requires
either the knowledge of the shear viscosity, or
a series of MD simulations with varying box-lengths.\cite{yeh_2004}

Recently, we have reported that direction-dependent
self-diffusion data, obtained from a single MD simulation 
with PBCs
based on
a specific orthorhombic
unit cell can be used to determine both the system size independent
true self-diffusion coefficient $D_0$ and the shear viscosity $\eta$.\cite{busch_2023c}
By performing MD simulations of orthorhombic systems 
with  ``magic'' box length ratios of 
$L_z/L_x\!=\!L_z/L_y\!=\!2.7933596497$,
due to a cancelling effect of
the hydrodynamic interactions,
the computed self-diffusion coefficients $D_x$ and $D_y$ in $x$- and $y$-direction 
are representing the {\em true} 
system size independent
self-diffusion coefficient $D_0\!=\!(D_x+D_y)/2$. At the same time
the shear viscosity can be determined from diffusion coefficients in $x$-,$y$- and $z$-direction using $\eta\!=\!k_\mathrm{B}T\cdot 8.1711245653/[3\pi L_z(D_{x}+D_{y}-2D_z)]$, 
where $k_\mathrm{B}$ denotes Boltzmann's constant
and $T$ represents the temperature.\cite{busch_2023c}
This approach was coined ``OrthoBoXY'', and is based on a
recently derived
extension of the Yeh-Hummer approach, allowing for a
quantitative
description of the anisotropy of the diffusion-tensor
of an isotropic fluid
caused by hydrodynamic interactions within an
orthorhombic periodic system.\cite{kikugawa_2015}
\begin{table*}[t]
\caption{\label{tab:madelung}
Madelung constant analogues
according to Equation \ref{eq:Ewald_aniso}, computed for orthorhombic MD simulation cells
with $L_x\!\neq\!L_y\!\neq\!L_z$, fulfilling the condition
$L_y/L_z\!=\! L_x/L_y$ for varying indicated sets of box-lengths ratios.
}
\setlength{\tabcolsep}{0.88cm}
        \centering              
\begin{tabular}{ccccc}
\hline\hline\\[-0.2em]
$L_y/L_z$ &
$L_x/L_z$ &
$\zeta_{xx}$ &
$\zeta_{yy}$ &
$\zeta_{zz}$ \\\hline\\[-0.6em]
0.95  & 0.9025 & 2.5828924663 & 2.828555577 & 3.096529075 \\
0.90  & 0.81   & 2.3170121640 & 2.800065379 & 3.378128871 \\
0.85  & 0.7225 & 2.0355569516 & 2.747235325 & 3.688644375 \\
0.80  & 0.64   & 1.7339175977 & 2.663352789 & 4.036025562 \\
0.75  & 0.5625 & 1.4069966828 & 2.538694622 & 4.429678724 \\
0.70  & 0.49   & 1.0490574329 & 2.359206961 & 4.880643368 \\
0.65  & 0.4225 & 0.6533320232 & 2.104440785 & 5.402004186 \\
0.60  & 0.36   & 0.2113689766 & 1.744178359 & 6.009538053 \\
0.57804765578&0.33413909235& 0& 1.541707906 & 6.308282188 \\[0.6em]
\hline\hline
\end{tabular}
\end{table*}

The use of this ``magic'' box shape has been deemed
particularly appealing, since
no further conversions are required to obtain the
true self-diffusion coefficient $D_0$ from the MD simulation data.
However, a slight drawback might be that 
the simulation box 
with box-length ratios $L_z/L_x\!=\!L_z/L_y\!=\!2.7933596497$,
is strongly elongated in the $z$-direction, 
which means that
rather large system sizes are required if the box-lengths in $x$- and
$y$-direction are meant to exceed a certain minimum threshold. 
In this contribution we therefore follow
a strategy to generate
 alternative box shapes,
offering 
the possibility to study 
boxes which resemble more closely
the popular cubic box.
In particular, we
explore 
box geometries with $L_x\!\neq\!L_y\!\neq\!L_z$,
which, however, obey the condition
$L_y/L_z\!=\! L_x/L_y$. Thus the box shape
can be controlled by systematically
varying the ratio $L_y/L_z$, while approaching
the form of a cubic box with
$L_y/L_z\!=\!1$ as a {limiting} case.
To study the efficiency of this 
approach, we determine the influence of the
box shape on the statistical
uncertainty of the computed values for $D_0$ and $\eta$.
In the following, we also derive expressions to
compute $D_0$ and $\eta$ purely from
direction-dependent self-diffusion data
under those conditions.

\section{Generalized OrthoBoXY-Approach for MD Simulations Using Arbitrary Orthorhombic Simulation Boxes}


For orthorhombic box geometries, the presence of
unequal box-lengths leads to 
different system size dependencies for each of the components 
$D_{ii}$ with ${i}\in\{x,y,z\}$ of the diffusion
tensor $\mathbf{D}$ such that 
the self-diffusion tensor becomes anisotropic
under PBCs
even for an isotropic fluid.\cite{kikugawa_2015a}
For  a quantitative description of this effect, 
Kikugawa et al.\ and 
others \cite{kikugawa_2015,voegele_2016} 
have followed the approach of Yeh and Hummer \cite{yeh_2004},
who had realized that a particle in a periodic system
experiences
hydrodynamic interactions not only with the solvent in 
its immediate surrounding, but also with its
periodic images, communicated via the solvent.
They have, 
based on the linearized Navier-Stokes 
equation for an incompressible fluid, and
the Kirkwood-Riseman theory of polymer diffusion \cite{kirkwood_1948},
obtained an expression for
the diffusion tensor modified for a periodic system with
\begin{equation}
\label{eq:yeh_hummer}
\mathbf{D}_\mathrm{PBC}
=
D_0\mathbf{1} + k_\mathrm{B}T \lim_{r\rightarrow 0}
\left[\mathbf{T}_\mathrm{PBC}(\mathbf{r})-
\mathbf{T}_\mathrm{0}(\mathbf{r})
\right]\;.
\end{equation}
Here 
$\mathbf{1}$ is the unity matrix, and
$\mathbf{T}_\mathrm{PBC}(\mathbf{r})$
and 
$\mathbf{T}_\mathrm{0}(\mathbf{r})$
are the Oseen mobility tensors for a periodic system
and an infinite nonperiodic system, respectively.\cite{yeh_2004}
$D_0$ denotes the {\em true} scalar diffusion coefficient within
an infinite system.
In order to bring equation \ref{eq:yeh_hummer} in a form, which
could be treated numerically, the technique of Ewald summation
adapted to hydrodynamic interactions
was employed.\cite{hasimoto_1959,beenacker_1986} The result is an
expression, describing the system size dependence
of self-diffusion coefficients from MD simulations
with PBCs, 
based on the effect of the hydrodynamic interactions
between particles in a periodic system.

Let us now assume that we have an 
orthorhombic simulation box
with $L_x\!\neq\!L_y\!\neq\!L_z$ in combination with PBCs,
and are performing an MD simulation of
an isotropic fluid.
For this situation, we can compute the 
direction-dependent 
self-diffusion coefficients $D_{\textrm{PBC},ii}$
based on Equation \ref{eq:yeh_hummer}
from the 
knowledge of the true self-diffusion coefficient $D_0$ 
by using
\begin{eqnarray}
\label{eq:Dpbc_aniso}
D_{\textrm{PBC},ii} &=& D_{0}
-
\frac{k_\mathrm{B}T\zeta_{ii}}{6\pi\eta L_{i}}\;,
\end{eqnarray}
where $\eta$ is the viscosity, and
$L_{i}$ are the individual box-lengths
of the orthorhombic unit cell.\cite{busch_2023c} 
The $\zeta_{ii}$ represent the
direction-dependent  Madelung constant analogues of the
orthorhombic lattice, which are calculated
by Ewald summation
 using \cite{kikugawa_2015,busch_2023c}
\begin{eqnarray}
\label{eq:Ewald_aniso}
\zeta_{ii}&=&-\frac{3}{2}\,L_{i}\cdot\Biggl\{
\,{\frac{1}{2}}\Biggl[\sum_{\mathbf{n}\neq 0}
\frac{\mathrm{erfc}({\alpha}\, n)}{n}\\\nonumber
&&+\frac{n_{i}^2}{n^2}
\left(
\frac{\mathrm{erfc}({\alpha}\, n)}{n}
+\frac{2{\alpha}}{\sqrt{\pi}} e^{-{\alpha}^2n^2}
\right)
\Biggr] \\\nonumber
&&+\frac{\pi}{V}
\Biggl[\sum_{\mathbf{k}\neq 0}
\frac{4 \,e^{-k^2/(4{\alpha}^2)}}{k^2}\\\nonumber
&&-\frac{k_{i}^2}{{\alpha}^2 k^2}
e^{-k^2/(4{\alpha}^2)}
\left( 
1+\frac{4{\alpha}^2}{k^2}
\right)
\Biggr] \\\nonumber
&&
-\frac{\pi}{{\alpha}^2V}
-\frac{\alpha}{\sqrt{\pi}}
\Biggr\}\;.
\end{eqnarray}
with $\mathbf{n}\!=\!(n_x,n_y,n_z)$,
and
$\mathbf{k}\!=\!(k_x,k_y,k_z)$
being real  and reciprocal lattice vectors with
$n_{i}\!=\!L_{i} m_{i}$ and
$k_{i}\!=\!2\pi\cdot m_{i}/L_{i}$,
based on integer numbers for $m_{i}$. We use
$n=|\mathbf{n}|$ and $k^2=|\mathbf{k}|^2$, while
$\alpha$ represents the Ewald convergence parameter.

From Equation \ref{eq:Dpbc_aniso} follows that by using
the difference between two system-size dependent self-diffusion coefficients
for two different directions $i$ and $j$,
we can obtain for any orthorhombic simulation box 
a term describing the viscosity $\eta$
as follows:
\begin{equation}
\label{eq:eta_ij}
\eta_{ij} = 
\frac{k_\mathrm{B}T\left(\zeta_{jj}/L_j-\zeta_{ii}/L_i\right)}
{6\pi \left(D_{\mathrm{PBC},ii}-D_{\mathrm{PBC},jj}\right)}\;.
\end{equation}
Here, the indices $ij$ indicate that the estimate of
the viscosity was obtained by employing 
directions $i$ and $j$. Of course,
sufficient sampling will lead to
$\eta\!=\!\eta_{xy}\!=\!\eta_{xz}\!=\!\eta_{yz}$. Consequently, the 
viscosity can then be computed by averaging over contributions from
all three different directions according to
\begin{equation}
\label{eq:eta}
\eta=\frac{1}{3}\left( \eta_{xy}+\eta_{xz}+\eta_{yz}\right)
\end{equation}
and the average true system size independent  self-diffusion coefficient $D_0$ can
be obtained from
\begin{eqnarray}
\label{eq:d0_new}
D_0 & = & \frac{1}{3}
\Biggl[
D_{\textrm{PBC},{xx}} + D_{\textrm{PBC},{yy}} +D_{\textrm{PBC},{zz}} 
 \\
&&
+ \frac{k_\mathrm{B}T}{6\pi\eta}
\left(
\frac{\zeta_{xx}}{L_x} + \frac{\zeta_{yy}}{L_y} + \frac{\zeta_{zz}}{L_z} 
\right)
\Biggr]\;.\nonumber
\end{eqnarray}
{ Note that $D_0$ is also directly available by combining Equations \ref{eq:Dpbc_aniso} and \ref{eq:eta_ij} without the need of 
involving both the viscosity $\eta$ and the temperature $T$, yielding
\begin{equation}
\label{eq:d0_nice}
D_0=D_{\mathrm{PBC},ii} +\frac{\frac{\zeta_{ii}}{L_i}}{\frac{\zeta_{jj}}{L_j}-\frac{\zeta_{kk}}{L_k}}
\left( D_{\mathrm{PBC},kk}-D_{\mathrm{PBC},jj} \right)
\end{equation}
with $i,j,k \in\{x,y,z\}$. The best estimate for $D_0$ 
following Equation \ref{eq:d0_nice}
is determined
by computing a weighted average of all three possible permutations of
of $x$, $y$, and $z$.
Keep in mind that the quantities $\eta_{ij}$ and 
$D_{\mathrm{PBC},ii}$ in Equations \ref{eq:eta} and \ref{eq:d0_new}
might have different uncertainties, such that their
averages should also be better computed as {\em weighted} averages.
To be able to make use of Equations \ref{eq:Dpbc_aniso},
\ref{eq:eta_ij}, \ref{eq:d0_new}, and \ref{eq:d0_nice},} we have
computed the Madelung constant analogues
in 
$x$-, $y$- and 
$z$-direction for various box-length ratios 
shown
in Table \ref{tab:madelung}.
The computations of Equation \ref{eq:Ewald_aniso} 
discussed above
were performed using double-precision floating point arithmetic,
and an Ewald convergence parameter of
${\alpha}\!=\!4/L_z$
with $m_{i}$ ranging between
$-m_\mathrm{max}\leq m_{i}\leq m_\mathrm{max}$ using
$m_\mathrm{max}\!=\!100$
for both the real and reciprocal lattice summation,
ensuring that all calculated
Madelung constant analogues $\zeta_{ii}$ 
shown in Table \ref{tab:madelung}
are converged.

\begin{table*}[t]
\caption{\label{tab:md_ortho}
Parameters describing the MD simulations
of 1536 TIP4P/2005 water molecules
using an orthorhombic unit cell with box-lengths ratios
$L_y/L_z\!=\! L_x/L_y$
 performed
under NVT conditions at a temperature of $T\!=\!298\,\mbox{K}$
and a density of
 $\rho\!=\!0.9972\,\text{g}\,\text{cm}^{-3}$.
Here
$L_x$, $L_y$, and $L_z$ representing the box lengths of the orthorhombic unit cell.  
The direction-dependent self-diffusion coefficients
$D_{\text{PBC},{ii}}$ are 
determined from the slope of the center-of-mass mean square displacement of the 
water molecules over a time interval between
$15\,\mbox{ps}$ and $200\,\mbox{ps}$.
}
\setlength{\tabcolsep}{0.32cm}
        \centering              
\begin{tabular}{ccccccc}
\hline\hline\\[-0.2em]
$L_y/L_z$ & 
$L_x/\text{nm}$ &
$L_y/\text{nm}$ &
$L_z/\text{nm}$ &
$D_{\mathrm{PBC},xx}  / 10^{-9}\mathrm{m}^2\mathrm{s}^{-1}$ &
$D_{\mathrm{PBC},yy}  / 10^{-9}\mathrm{m}^2\mathrm{s}^{-1}$ &
$D_{\mathrm{PBC},zz}  / 10^{-9}\mathrm{m}^2\mathrm{s}^{-1}$ 
\\\hline\\
0.95 & 3.40592 & 3.58517 & 3.77387 & $2.1032\pm0.0050$ & $2.1004\pm0.0046$ & $2.0911\pm0.0041$\\
0.90 & 3.22666 & 3.58517 & 3.98353 & $2.1145\pm0.0046$ & $2.1013\pm0.0045$ & $2.0777\pm0.0044$\\
0.85 & 3.04740 & 3.58517 & 4.21785 & $2.1330\pm0.0050$ & $2.1022\pm0.0045$ & $2.0787\pm0.0044$\\
0.80 & 2.86814 & 3.58517 & 4.48147 & $2.1519\pm0.0050$ & $2.1019\pm0.0044$ & $2.0793\pm0.0040$\\
0.75 & 2.68888 & 3.58517 & 4.78023 & $2.1618\pm0.0053$ & $2.1177\pm0.0048$ & $2.0662\pm0.0039$\\
0.70 & 2.50962 & 3.58517 & 5.12168 & $2.1957\pm0.0061$ & $2.1295\pm0.0057$ & $2.0545\pm0.0040$\\
0.65 & 2.33036 & 3.58517 & 5.51565 & $2.2284\pm0.0070$ & $2.1478\pm0.0061$ & $2.0564\pm0.0038$\\
0.60 & 2.15110 & 3.58517 & 5.97529 & $2.2740\pm0.0071$ & $2.1763\pm0.0071$ & $2.0508\pm0.0039$\\
$0.578047\ldots$ & 2.07240 & 3.58517 & 6.20221 & $2.2964\pm0.0077$ & $2.1841\pm0.0074$ & $2.0484\pm0.0041$\\[0.6em]
\hline\hline
\end{tabular}
\end{table*}

Here, we consider systems with an orthorhombic unit cell
with $L_x\!\neq\!L_y\!\neq\!L_z$. In particular, 
we investigate unit cells where the box-length ratios are connected
with respect to one another
via $L_y/L_z\!=\! L_x/L_y$,
leading to the relation $L_x/L_z=(L_y/L_z)^2$. This allows us to 
deviate from a cubic box geometry in a systematic
fashion by varying the ratio $L_y/L_z$
as shown in Table \ref{tab:madelung}. Here
$L_y/L_z\!=\!1$ represents the limiting value for cubic boxes.

Note that
under this condition, there exists at least one other set of
``magic'' box-length ratios with
$L_y/L_z\!=\!0.57804765578$ and
$L_x/L_z\!=\!0.33413909235$, leading to a Madelung constant analog in $x$-direction
$\zeta_{xx}\!=\!0$. This indicates that for
such an MD box, the self-diffusion in
$x$-direction is representing the true system size independent
self-diffusion coefficient 
\begin{equation}
\label{eq:d0_magic}
D_0=D_{\mathrm{PBC,} xx}\;,
\end{equation}
and the shear viscosity can be determined using
the self-diffusion data in $y$- and $z$-direction:
\begin{equation}
\label{eq:eta_magic}
\eta = 
\frac{k_\mathrm{B}T}
{12\pi}\cdot
\left[
\frac{\zeta_{yy}/L_y}
{D_0-D_{\mathrm{PBC},yy}}
+
\frac{\zeta_{zz}/L_z}
{D_0-D_{\mathrm{PBC},zz}}
\right]\;
\end{equation}
with
$\zeta_{yy}\!=\!1.541707906$ and
$\zeta_{zz}\!=\!6.308282188$ (see also Table \ref{tab:madelung}).
Since the computation of the $\zeta_{ii}$ has been performed numerically,
we have determined
$\zeta_{xx}\!<\! 10^{-10}$ using the ``magical'' box geometry indicated above.
{ Note, however, that also for $\eta$ in Equation \ref{eq:eta_magic}, a weighted
average might be in most cases a more appropriate choice.}

\section{Molecular Dynamics Simulations}

To test the 
generalized
``OrthoBoXY''
approach outlined in the previous section, MD  simulations of
TIP4P/2005 model 
water \cite{abascal_2005}
were  carried out. TIP4P/2005 has been demonstrated to accurately describe the
properties of water compared to other
simple rigid nonpolarizable water models.\cite{vega_2011}
Simulations were performed
at a temperature of $T\!=\!298\,\mbox{K}$
under NVT conditions
at a density of
 $\rho\!=\!0.9972\,\text{g}\,\text{cm}^{-3}$.
 Systems containing 1536 water molecules are simulated
 using various orthorhombic box geometries 
 fulfilling the condition $L_y/L_z\!=\! L_x/L_y$.
 The studied box geometries are summarized in Table \ref{tab:madelung}.
 MD simulations of 160\,ns length each were performed
 using \textsc{Gromacs} 5.0.6.\cite{gromacs4,gromacs3}
The integration time step for all simulations was $2\,\mbox{fs}$.
The temperature of the simulated systems was controlled employing the
Nos\'e-Hoover thermostat~\cite{Nose:1984,Hoover:1985}
with
a coupling time $\tau_T\!=\!1.0\,\mbox{ps}$.
Both, the Lennard-Jones and electrostatic interactions were treated by smooth 
particle mesh Ewald summation.\cite{Essmann:1995,wennberg_2013,wennberg_2015}
The Ewald convergence parameter 
was set to a relative accuracy of the Ewald sum of $10^{-5}$ 
for the Coulomb-interaction and $10^{-3}$ for the LJ-interaction. 
All bond lengths were kept fixed during the simulation run and
distance constraints were solved by means of 
the SETTLE procedure. \cite{miyamoto_1992}
The simulations were carried out in 
$n_\mathrm{W}\!=\!320$ subsequent segments of 
$\tau_\mathrm{W}\!=\!500\,\mbox{ps}$ length,
resulting in total simulation times 
of $160\,\mbox{ns}$ each. 
For each simulation segment, 2500 frames were stored with
a time interval of $0.2\,\mbox{ps}$ between consecutive frames.
All reported properties were 
then calculated for 
each of the segments separately to be able to estimate the 
uncertainty
using standard statistical analysis procedures.\cite{allentildesley,numrecipes}
Here the variance $\sigma_X$ of a computed property $X$ is estimated via
\begin{equation}
\sigma^2_X= \langle X^2\rangle - \langle X\rangle^2\;,
\end{equation}
where $\langle\ldots\rangle$ indicates averaging over $n_\mathrm{W}$
simulation run segments, 
and its uncertanity is determined via 
\begin{equation}
\hat{\sigma}_X= 
\frac{\sigma_X}{\sqrt{n_\mathrm{W}}}\;.
\end{equation}

All simulation boxes listed in Table \ref{tab:md_ortho} were
created starting from a single
simulation box with 
$L_x\!=\!L_y\!=\!2.48582\,\mbox{nm}$ and
$L_z\!=\!7.45747\,\mbox{nm}$, 
containing 1536 TIP4P/2005 water
molecules at a density of 
$\rho\!=\!0.9972\,\text{g}\,\text{cm}^{-3}$,
which is available from our GitHub repository.\cite{github-orthoboxy}
The boxes were then morphed into their final form
in a volume-preserving fashion 
by using short nonequilibrium MD simulation runs of
$200\,\mbox{ps}$ length by employing GROMACS' ``deform'' feature.\cite{gromacs_doc}
The prepared systems were then equilibrated under NVT conditions for another 
$500\,\mbox{ps}$.

\section{Results and Discussion}

\begin{figure*}[t]
        \centering
        \includegraphics[width=0.4\textwidth]{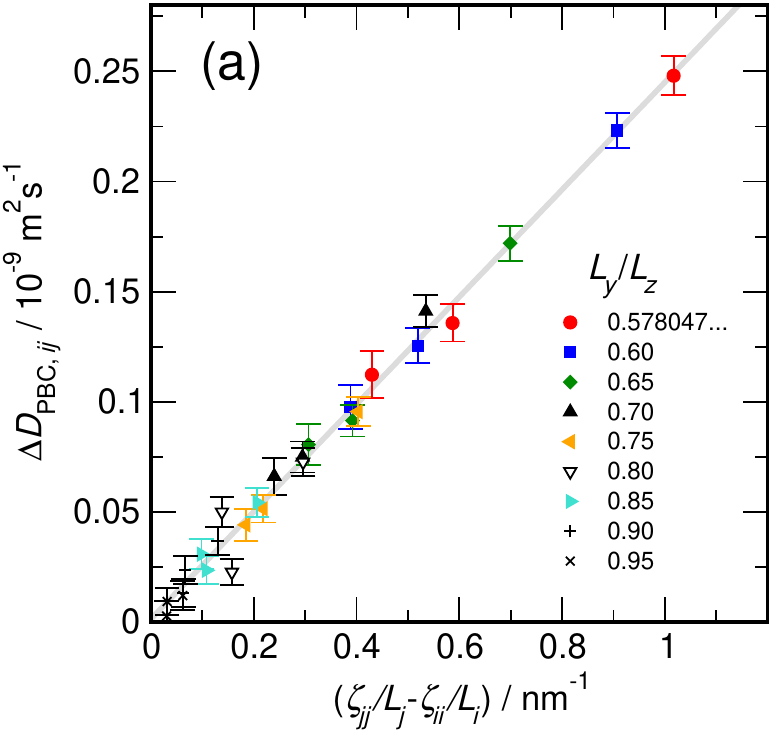}  
        \hspace*{2em}            
        \includegraphics[width=0.385\textwidth]{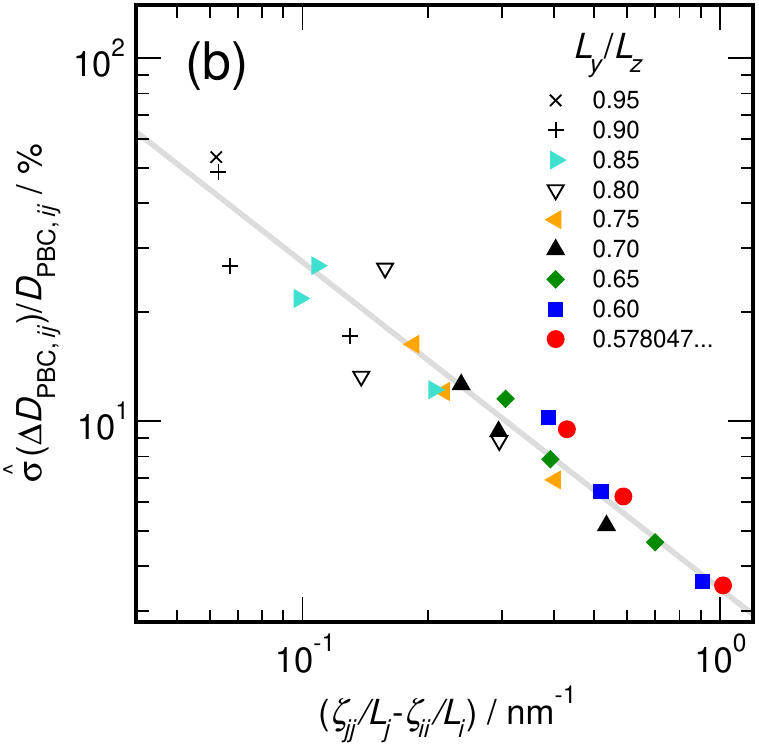}              
        \caption{\label{fig:dplot} 
        Analysis of the direction-dependent self-diffusion data
        according to Equation \ref{eq:eta_ij}.
        a)
        Differences of the 
        self-diffusion coefficients in $i$- and $j$-direction
        $\Delta D_{\mathrm{PBC},ij}\!=\!D_{\mathrm{PBC},ii}-D_{\mathrm{PBC},jj}$
        with ${i,j}\in\{x,y,z\}$
        vs. $\zeta_{jj}/L_j-\zeta_{ii}/L_i$ according to 
        Equation \ref{eq:eta_ij}
        for TIP4P/2005 water at $298\,\mbox{K}$ determined from
        MD simulations employing orthorhombic simulation boxes
        with varying box-lengths ratios $L_y/L_z\!=\! L_x/L_y$.
        The grey solid line represent a linear fit of the data according to
        Equation \ref{eq:eta_ij}, resulting in a viscosity
        of $\eta\!=\!(0.8957\pm 0.0178)\,\mbox{mPa}\,\mbox{s}$.
        b)
        Log-log plot of
        the relative error of the computed differences of
        the self diffusion coefficients
        $\hat{\sigma}(\Delta D_{\mathrm{PBC},ij})/\Delta D_{\mathrm{PBC},ij}$
        given in percent, vs. $\zeta_{jj}/L_j-\zeta_{ii}/L_i$.
        The solid grey line represents a fitted scaling of the errors
        according to $(\zeta_{jj}/L_j-\zeta_{ii}/L_i)^{\beta}$ with an
        exponent $\beta\!=\!-0.9$.
        }
\end{figure*}
\begin{figure*}[t]
        \centering
        \includegraphics[width=0.375\textwidth]{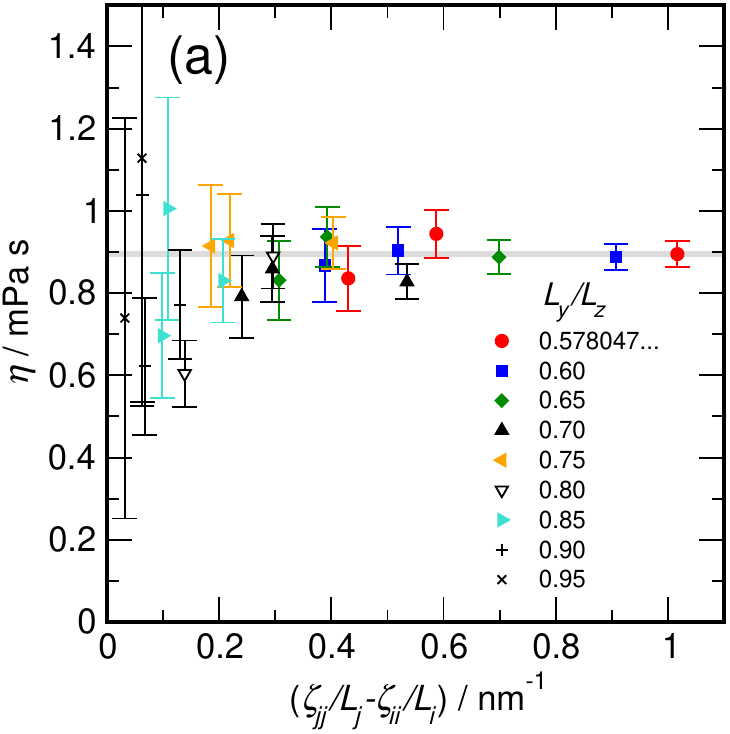}  
        \hspace*{2em}            
        \includegraphics[width=0.395\textwidth]{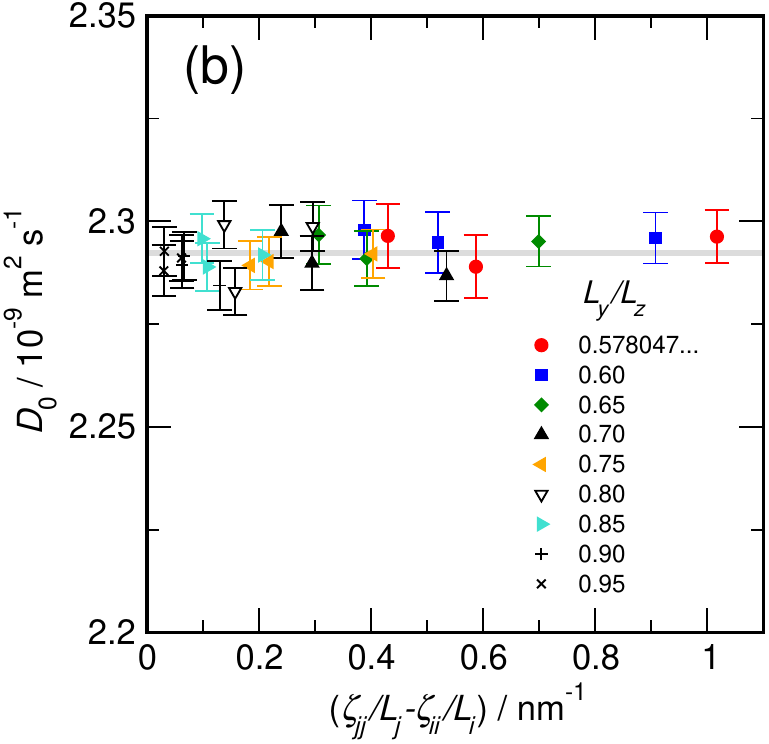}              
        \caption{\label{fig:d0etaplot} 
        Shear viscositiies $\eta$ and 
        system size independent true self-diffusion coefficients $D_0$
        obtained for
        TIP4P/2005 water at $298\,\mbox{K}$, determined from
        MD simulations employing orthorhombic simulation boxes
        with varying box-lengths ratios $L_y/L_z\!=\! L_x/L_y$.
        a)
        Shear viscosities 
        computed according to 
        Equation \ref{eq:eta_ij} using self-diffusion coefficients
        listed in Tables \ref{tab:madelung} and \ref{tab:md_ortho}.
        The grey solid line represents a viscosity of 
        $\eta\!=\!0.8957\,\mbox{mPa}\,\mbox{s}$,
        representing 
        a linear fit of the data 
        hown in Figure \ref{fig:dplot} according to 
        Equation \ref{eq:eta_ij}.
        b)
        True self-diffusion coefficients
        $D_0$ obtained according to Equation \ref{eq:Dpbc_aniso}
        from data listed 
        in Tables \ref{tab:madelung} and \ref{tab:md_ortho}.
        using the average viscosity 
        $\eta\!=\!(0.8957\pm0.0178)\,\mbox{mPa}\,\mbox{s}$.
        The grey solid line represents the average value
        $D_0\!=\!2.2923\times 10^{-9}\,\mbox{m}^2\,\mbox{s}^{-1}$.
        }
\end{figure*}

Self-diffusion coefficients were computed 
from the slope of the center-of-mass mean square displacement of the 
water molecules using the Einstein formula \cite{allentildesley} according to
\begin{equation}
D_\textrm{PBC}
=\frac{1}{6}
\frac{\partial}{\partial t}
\lim_{t\rightarrow\infty}
\left<
|\mathbf{r}(0)
-
\mathbf{r}(t)
|^2
\right>\;,
\end{equation}
and
\begin{equation}
D_{\textrm{PBC},{ii}}
=\frac{1}{2}
\frac{\partial}{\partial t}
\lim_{t\rightarrow\infty}
\left<
|r_{i}(0)
-
r_{i}(t)
|^2
\right>\;,
\end{equation}
where $\mathbf{r}(t)=[r_x(t),r_y(t),r_z(t)]$ represent the position of the center of mass
of a water molecule at time $t$ and the $r_{i}(t)$ are its respective components
in $x$-, $y$-, and $z$-direction.
All computed self-diffusion coefficients 
shown in Table \ref{tab:md_ortho}  
were determined from the slope of the mean square
displacement of the water molecules
fitted to time intervals between $15\,\mbox{ps}$ and $200\,\mbox{ps}$.

Table \ref{tab:md_ortho} contains
direction-dependent self-diffusion coefficients $D_{\textrm{PBC},{ii}}$
obtained
from MD simulations using
orthorhombic unit cells with
$L_y/L_z=L_x/L_y$ for various ratios
$L_y/L_z$. Here a decreasing ratio
$L_y/L_z$ indicates an increasing deviation from the cubic form.
If the hydrodynamic interaction approach 
towards self-diffusion
according to 
Equation \ref{eq:yeh_hummer} holds, 
we expect a linear relation 
between the difference between
self-diffusion coefficients in different directions
$i$ and $j$ with
$\Delta D_{\mathrm{PBC},ij}\!=\!D_{\mathrm{PBC},ii}-D_{\mathrm{PBC},jj}$
and the difference between the 
Madelung constant analogue weighted with inverse box-lengths
$\zeta_{jj}/L_j-\zeta_{ii}/L_i$
according to Equation \ref{eq:eta_ij}.
Figure \ref{fig:dplot}a demonstrates, that this linear relationship is
excellently fulfilled. 
A linear fit including all $9\times 3\!=\!27$ 
direction-dependent data points
passes almost perfectly
through the origin with 
$\Delta D_{\mathrm{PBC},ij}(0)=(0.0013\pm 0.0020)\times 10^{-9}\,\mbox{m}^2\,\mbox{s}^{-1}$,
while the slope of the fitted solid grey
line in \ref{fig:dplot}a yields a viscosity
of $\eta\!=\!(0.8957\pm 0.0178)\,\mbox{mPa}\,\mbox{s}$, which is consistent
with the value 
$\eta\!=\!(0.900\pm 0.051)\,\mbox{mPa}\,\mbox{s}$ 
obtained 
for TIP4P/2005 under the same conditions (temperature and density)
in Ref.~\cite{busch_2023c}
It is evident, however, that for near-cubic box geometries
with $L_y/L_z\geq0.95$ the error 
and the magnitude of $\Delta D_{\mathrm{PBC},ij}$ have almost the same size,
thus rendering those systems an extremely unreliable source for
determining viscosity data. This behavior is a consequence of the
little variation found for the size of the computed uncertainties
of the direction-dependent diffusion coefficients,
as can be seen from the data shown in Table \ref{tab:md_ortho}.  
This is leading to a strong increase of the
relative error of $\Delta D_{\mathrm{PBC},ij}$ for
$L_y/L_z\rightarrow 1$.
This behavior is  perfectly demonstrated 
in Figure \ref{fig:dplot}b in the form of a
log-log plot of
the relative error of the computed differences of
the self diffusion coefficients
$\hat{\sigma}(\Delta D_{\mathrm{PBC},ij})/\Delta D_{\mathrm{PBC},ij}$
vs. $\zeta_{jj}/L_j-\zeta_{ii}/L_i$.
Given the linear dependence
of $\Delta D_{\mathrm{PBC},ij}$ vs. $\zeta_{jj}/L_j-\zeta_{ii}/L_i$
as shown in Figure \ref{fig:dplot}a,
a constant size of $\hat{\sigma}(\Delta D_{\mathrm{PBC},ij})$
would lead to an inverse proportional behavior according to
\begin{equation}
\frac{\hat{\sigma}(\Delta D_{\mathrm{PBC},ij})}{\Delta D_{\mathrm{PBC},ij}}
\propto\left(\frac{\zeta_{jj}}{L_j}-\frac{\zeta_{ii}}{L_i}\right)^\beta
\end{equation}
with an exponent $\beta\!=\!-1$.
Due to small, albeit significant
variation of $\hat{\sigma}(\Delta D_{\mathrm{PBC},ij})$
as a function of $L_y/L_z$, however, the
solid grey line 
in Figure \ref{fig:dplot} indicates a 
scaling with a slightly smaller exponent $\beta\!=\!-0.9$.

According to Equation \ref{eq:eta_ij}, we can compute
an estimate of the viscosity $\eta_{ij}$
for each difference
in diffusion coefficients $\Delta D_{\mathrm{PBC},ij}$. These
viscosity estimates are shown in Figure \ref{fig:d0etaplot}a
as a function of $\zeta_{jj}/L_j-\zeta_{ii}/L_i$.
The predictive value of the most anisotropic systems 
is very good, the quality of the estimate, however, drops significantly
for $L_y/L_z\rightarrow 1$, since
$\hat{\sigma}(\eta_{ij})/\eta_{ij}\!\approx\!\hat{\sigma}(\Delta D_{\mathrm{PBC},ij})/\Delta D_{\mathrm{PBC},ij}$. In Figure \ref{fig:d0etaplot}b, the computed
estimates for the true self-diffusion coefficients $D_0$ according to
Equation \ref{eq:Dpbc_aniso} are shown. To compute $D_0$, we have used
the average viscosity of $\eta\!=\!(0.8957\pm0.0178)\,\mbox{mPa}\,\mbox{s}$.
Note that Equation \ref{eq:Dpbc_aniso} holds up extremely well, leading
to a very small statistical
variation of the computed average for $D_0$ with no recognizable trend
as a function of $L_y/L_z$,
including the simulations with near-cubic box shapes.
Averaging those values leads to
$D_0=(2.2923\pm0.0008)\times 10^{-9}\,\mbox{m}^2\,\mbox{s}^{-1}$. The
accuracy suggested by the small
statistical error, however, might be misleading, since it
does not properly account for the uncertainty of the 
average viscosity, which
would shift the whole ensemble of data points in Figure
\ref{fig:d0etaplot}b simultaneously up or down.

Next, we would like to compute both $D_0$ and $\eta$ from the system
with ``magical'' box length ratios
of
$L_y/L_z\!=\!0.57804765578$ and
$L_x/L_z\!=\!0.33413909235$.
Here, according to Equation \ref{eq:d0_magic} 
$D_0=D_{\mathrm{PBC},xx}=(2.2964\pm0.0077)\times 10^{-9}\,\mbox{m}^2\,\mbox{s}^{-1}$,
which is in excellent agreement with the estimate including all box shapes.
The weighted average of the viscosity computed according to
Equation \ref{eq:eta_magic} is
{$\eta\!=\!(0.8869\pm0.0415)\,\mbox{mPa}\,\mbox{s}$},
which is { slightly}
smaller than the estimate from the slope 
of the data in Figure \ref{fig:dplot}a, 
albeit within the range of the statistical uncertainty.
Note that in Ref. \cite{busch_2023c}, the estimate for $D_0$ 
and the 
viscosity for a system of the same 
{ number of molecules at the same density}, but for a 
shorter simulation
of just $10\,\mbox{ns}$ length (with $n_\mathrm{W}\!=\!20$)
{was obtained as}
$D_0=(2.283\pm0.027)\times 10^{-9}\,\mbox{m}^2\,\mbox{s}^{-1}$
and
$\eta\!=\!(0.853\pm0.084)\,\mbox{mPa}\,\mbox{s}$.
Given the 16 times longer simulation runs,
we would expect a four times smaller statistical uncertainty, which
coincides rather well with the computed uncertainty for $D_0$.
The fact that the statistical uncertainty of $\eta$ is just
about one-half of the value reported in Ref. \cite{busch_2023c},
suggests the ``magical'' simulation-box procedure proposed here is less effective
in estimating the viscosity than the original
``OrthoBoXY''-method proposed in Ref. \cite{busch_2023c}.

Note that the computed viscosities are all lying very
close to the experimental
value for ordinary water between $0.892\,\mbox{mPa\,s}$ and
$0.893\,\mbox{mPa\,s}$ 
at $298.15\,\mbox{K}$, reported by Harris and Woolf \cite{harris_2004,harris_2004c}
(when using their corrected data tables listed in Ref.\cite{harris_2004c}),
while the self-diffusion coefficients
almost perfectly match the experimental value of
Krynicki et al.\cite{krynicki_1978}
with $2.30\times 10^{-9}\,\text{m}^2\,\text{s}^{-1}$ 
at $298.2\,\mbox{K}$,
and of Mills \cite{mills_1973} with
$2.299\times 10^{-9}\,\text{m}^2\,\text{s}^{-1}$
at $25^\circ\mbox{C}$.

Finally, it should not be left unnoticed that the data shown in 
Figures \ref{fig:dplot}a and \ref{fig:d0etaplot}b provide
excellent evidence for the validity of the 
hydrodynamic-interaction-based approach for correcting self-diffusion coefficients
for non-cubic geometries.

\section{Conclusion}

We have derived equations representing a
generalized ``OrthoBoXY'' procedure, which can be
applied to MD simulations of any orthorhombic box geometry for
determining both the system size independent 
true self-diffusion coefficient $D_0$ and
the shear viscosity $\eta$.
We have tested this approach by using
NVT MD simulations of systems containing 1536 TIP4P/2005 water 
molecules
at a density
of $\rho\!=\!0.9972\,\text{g}\,\text{cm}^{-3}$
and a temperature of 
$T\!=\!298\,\mbox{K}$ using
varying box geometries. These systems obey 
the condition $L_y/L_z\!=\! L_x/L_y$,
while the ratio $L_y/L_z$ was systematically varied.
In particular,
we have explored the feasibility of employing
box shapes more akin to the cubic form.

We have demonstrated, that
we are indeed able to determine the {\em true} 
self-diffusion coefficient $D_0$ for TIP4P/2005 water without prior knowledge
of the shear viscosity 
from single MD simulation runs
using this generalized approach, 
similar
to what we have achieved in our previous paper
using simulation boxes with
$L_z/L_x\!=\!L_z/L_y\!=\!2.7933596497$.\cite{busch_2023c} 
The computed values for $D_0$ agree well with the values
determined from MD simulations employing 
both orthorhombic and
cubic unit cells 
discussed in Ref.\cite{busch_2023c}. 
Moreover, both the computed self-diffusion coefficient and shear viscosity 
agree nearly quantitatively with the experimentally observed
data for water at $298\,\mbox{K}$.

However, the idea, to use box shapes more akin to the cubic form 
turns out to be only partially practical, since the small differences 
in direction-dependent self-diffusion 
coefficients 
observed for near-cubic geometries
lead to unsustainably large
relative uncertainties for the computed viscosities. Large
differences between box-length, are instead the preferred choice.
This 
leads us to the conclusion, that the original ``OrthoBoXY''-approach
outlined in Ref.\cite{busch_2023c} 
has to be considered already quite efficient.

Instead, another ``magical'' set of box-length ratios has been
discovered
 with $L_y/L_z\!=\!0.57804765578$ and
$L_x/L_z\!=\!0.33413909235$, where the self-diffusion coefficient
in $x$-direction becomes system sized independent,
such that $D_0\!=\!D_{\mathrm{PBC,} xx}$. An expression 
for determining the viscosity $\eta$, employing  
$D_{\mathrm{PBC,} yy}$ and 
$D_{\mathrm{PBC,} zz}$, is given by Equation \ref{eq:eta_magic}.
However, from the standpoint of box anisotropy, this
box shape is deemed
less preferable to the original ``OrthoBoXY''-approach,
since the smallest box-length is about $20$ per cent smaller
than the smallest box length of a system of the
same size using the original ``OrthoBoXY''-approach
under the same conditions. Moreover, the simulations indicate,
that also uncertainty  of the predicted viscosity is slightly
worse than that of the original ``OrthoBoXY''-procedure.

\section*{Acknowledgements}

We thank the computer center at the University of 
Rostock (ITMZ) for providing and maintaining computational resources. 

\section*{Data Availability Statement}

The codes of 
\href{https://www.gromacs.org}{GROMACS} 
and
\href{https://www.moscitomd.org}{MOSCITO} 
are
freely available. 
Input parameter and topology files for the MD simulations and the 
code for computing the Madelung constant analogues for cubic and orthorhombic
lattices can be downloaded from GitHub  via
\href{https://github.com/Paschek-Lab/OrthoBoXY/}{https://github.com/Paschek-Lab/OrthoBoXY/}


\end{document}